# Strain-Switchable Field-Induced Superconductivity

Joshua J. Sanchez[1,3,*], Gilberto Fabbris[2], Yongseong Choi[2], Jonathan M. DeStefano[3], Elliott Rosenberg[3], Yue Shi[3], Paul Malinowski[3,4], Yina Huang[5], Igor I. Mazin[6], Jong-Woo Kim[2], Jiun-Haw Chu[3,*], Philip Ryan[2,*]

[1]Department of Physics, Massachusetts Institute of Technology, Cambridge, MA 02139, USA

[2]Advanced Photon Source, Argonne National Laboratories, Lemont, IL, USA.

[3]Department of Physics, University of Washington, Seattle, WA, USA.

[4]Department of Physics, Cornell University, Ithaca, NY 14853, USA.

[5]School of Science, Zhejiang University of Science and Technology, Hangzhou 310023, China

[6]Department of Physics and Astronomy and Quantum Science and Engineering Center, George Mason University, Fairfax, Virginia 22030, USA

*Corresponding authors: jjsanchez.physics@gmail.com, jhchu@uw.edu, pryan@anl.gov

**Abstract:** Field-induced superconductivity is a rare phenomenon where an applied magnetic field enhances or induces superconductivity. This fascinating effect arises from a complex interplay between magnetism and superconductivity, and it offers the tantalizing technological possibility of an infinite magnetoresistance superconducting spin valve. Here, we demonstrate field-induced superconductivity at a record-high temperature of T=9K in two samples of the ferromagnetic superconductor Eu(Fe$_{0.88}$Co$_{0.12}$)$_2$As$_2$. We combine tunable uniaxial stress and applied magnetic field to shift the temperature range of the zero-resistance state between 4K and 10K. We use x-ray diffraction and spectroscopy measurements under stress and field to demonstrate that stress tuning of the nematic order and field tuning of the ferromagnetism act as independent tuning knobs of the superconductivity. Finally, DFT calculations and analysis of the Eu dipole field reveal the electromagnetic mechanism of the field-induced superconductivity.

**Introduction**

The switching between distinct electronic phases in quantum materials by external tuning parameters is a central focus of condensed matter physics, both to study how competing orders interact and with the goal of technological development (*1*). One rich research area is tuning systems with both ferromagnetism and superconductivity. The interaction of the antagonistic phases leads to unusual phenomena, such as spontaneous magnetic vortices (*2*, *3*) and spin-polarized supercurrents (*4*), the latter of which hold promise for superconducting spintronics technologies and energy-efficient data storage. Much attention has focused on superconducting spin valves, i.e. thin film heterostructures with ferromagnetic layers surrounding a superconducting layer (*5*). An applied magnetic field switches the sandwiching ferromagnetic layers between parallel and antiparallel alignment, which strongly tunes the magnetic pairbreaking effect and effectively turns the superconductivity on and off. This enables the ultimate switchability of magneto-transport, between a resistive and zero-resistance state, thus achieving infinite magnetoresistance and the possibility of low energy dissipation computation technologies (*4*).

Infinite magnetoresistance occurs not only in artificial heterostructures, but also in a handful of single crystal materials exhibiting field-induced superconductivity, including several Eu and U-based superconductors (*6–10*) and organic superconductors (*11*, *12*). In these systems as well as in thin-film superconducting spin valves, the zero-resistance temperature $T_0$ is often below 1K, up to 4K for $UTe_2$ under pressure (*13*), limiting their practical application. The current record-holder for highest field-induced superconductivity temperature is in the chemically-doped Eu-based iron pnictide superconductor, $EuFe_2As_2$. Like other iron pnictide superconductors, $EuFe_2As_2$ exhibits an electronic nematic transition which creates orthorhombic structural twin domains. The suppression of nematicity by chemical doping results in the emergence of superconductivity, with an onset temperature $T_{SC}$ reaching 18K-30K at optimal doping (*14–17*). Meanwhile, in optimal doped materials the Eu moments order ferromagnetically along the c-axis, with $T_{FM}$=16K-20K (*18–22*). The similar ordering temperatures of the two antagonistic phases implies a potentially strong competition between them. Indeed, for Co- and Rh-doped samples a large reentrant resistivity appears below $T_{FM}$ as the Eu magnetic flux disrupts the nascent superconductivity, pushing $T_0$ far below $T_{SC}$. Unexpectedly, applying a small in-plane magnetic field ($\mu_0 H$ <0.5T) to these materials raises $T_0$ from ~5K to ~6-7K (*23*, *24*). Thus far, the mechanism of this field-induced superconductivity has not been determined, nor has the effect been optimized to enhance $T_0$ to its limit.

In this work, we demonstrate field-induced superconductivity in 12% Co-doped $EuFe_2As_2$ at $T_0$=9K, which can be enhanced up to at least 10K or suppressed to at least 4K using in-situ applied uniaxial stress. To our knowledge, this is the highest reported temperature of magnetic-field-induced superconductivity in any material. Doped $EuFe_2As_2$ exists as a natural-grown atomic limit of the thin film superconducting spin valve architecture, with alternating ferromagnetic Eu and superconducting/nematic FeAs layers (Fig.1A). We combine synchrotron x-ray techniques and transport measurements to reveal that strain-tuning of the nematicity and field-tuning of the Eu moments act as *independent* tuning knobs of the superconductivity (Fig.1B,C). Indeed, doped $EuFe_2As_2$ acts as a strain-switchable superconducting spin valve, which has potential both for spintronics applications and more fundamental investigations of ferromagnetic superconductors. Finally, we combine DFT calculations and analysis of the Eu dipole field to ascertain the origin of the field-induced superconductivity; in short, the directional anisotropy of the upper critical field $H_{c2}$ enables the in-plane reorientation of the Eu moments to boost $T_0$. In the Discussion, we consider how this novel mechanism could be realized in other systems, including in 2D systems and at even higher temperatures.

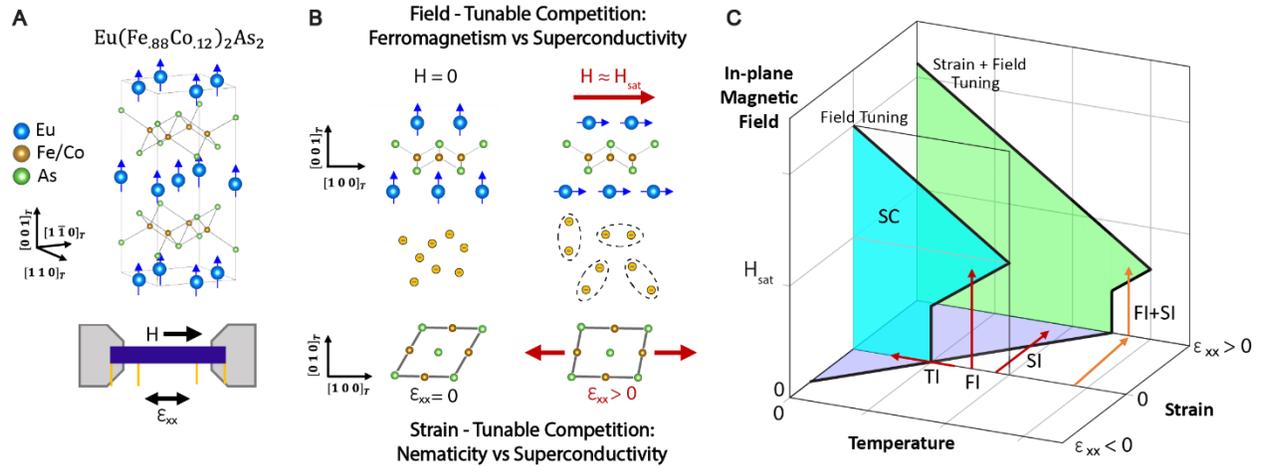

**Figure 1. Three Doors into Superconductivity.** (**A**) $\mathrm{Eu(Fe_{.88}Co_{.12})_2As_2}$ consists of stacked planes of Eu and FeAs layers, with the former exhibiting ferromagnetism (FM; $T_{FM}$=17K) and the latter hosting both nematicity (N, $T_S$=66K) and superconductivity (SC, $T_{SC}$=19K). The FM/SC phase competition prevents a zero-resistance state from being reached until the lower temperature $T_0$, below which the three phases simultaneously coexist. (**B,C**) Three doors (i.e. tuning parameters) lead to the zero-resistance state (shaded areas of the phase diagram). One door is field-induced superconductivity (FI, cyan). Applying a small in-plane magnetic field ($H_{sat}$) reorients the Eu moments and reduces the magnetic flux through the FeAs layers, enhancing superconductivity and boosting $T_0$. A second door is strain-induced superconductivity (SI, lavender). As in other iron-pnictide superconductors, the N/SC phase competition enables an effective strain-tuning of superconductivity via strain-tuning the lattice-coupled nematic order. Here, tensile (compressive) stress along the FeAs bonding direction enhances (suppresses) superconductivity and increases (decreases) $T_0$. The third door is simply tuning the temperature (TI) to cross the externally-tuned value of $T_0$. With combined strain and field tuning, $T_0$ can in principle take any value between T=0 and T=$T_{FM}$ (green).

**Strain-switchable field-induced superconductivity**

Single crystal samples of 12% Co-doped EuFe$_2$As$_2$ were grown using Sn flux (see Methods). We found that using a (Fe,Co)-rich, nonstoichiometric growth composition yielded samples with increased superconducting transition temperatures relative to stoichiometric-grown samples (*23*) (Methods, Supp.Fig.S1). Samples 1 and 2 were selected from different growth batches and were prepared identically as matchsticks to measure the inline resistivity $\rho_{xx}$ (Fig.1A). To better compare the field and strain tuning of the resistivity, we present all transport data normalized to the zero-field freestanding resistivity at T=25K, with $\rho/\rho_0 = \rho_{xx}(T, B, \varepsilon_{xx})/\rho_{xx}(25\,K, 0, 0)$.

In the freestanding state, sample 1 was cooled through the superconducting (T$_{SC}$=19K) and ferromagnetic (T$_{FM}$=17.2K) transitions under zero field (Fig.2, black), reaching $\rho/\rho_0 = 0$ at T$_0$=7.5K. Temperature sweeps were repeated with fixed magnetic field applied either in-plane (Fig.2, red) or out of plane (Fig.2, blue). An out of plane field is found to only increase the resistivity, while only lowering the value of T$_0$. In sharp contrast, an in-plane field is far more detrimental to superconductivity between T$_{SC}$ and T$_{FM}$, but zero resistance is reached at an enhanced value of T$_0$=9.0K for μ$_0$H=0.2T. Thus, we demonstrate field-induced superconductivity. It is striking that the superconductivity shows a different preference for in vs out of plane field above and below T$_{FM}$; we will return to this point in the section that discusses the mechanism of field-induced superconductivity.

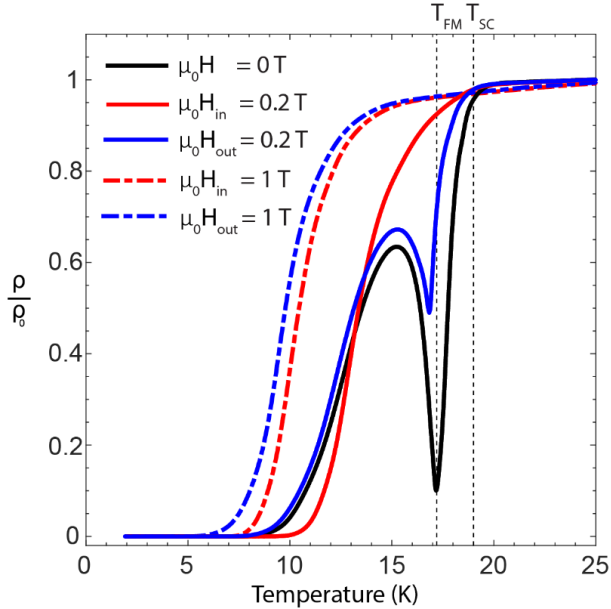

**Figure 2. Zero-strain field-induced superconductivity.** Sample 1 resistivity vs temperature for zero applied field (black) and μ$_0$H=0.2T (solid line) and 1T (dashed line) applied in-plane (red) and out of plane (blue). For μ$_0$H=0.2T applied in-plane, the zero-resistivity temperature rises from T$_0$=7.5K to 9.0K.

Figure 3A shows $\rho/\rho_0$ vs temperature at fixed in-plane field (μ₀H=0 T and μ₀H=1 T) and $\rho/\rho_0$ vs field at fixed temperature for sample 1. For T > T$_{FM}$, an applied field up to 1 T acts only to increase the resistivity. For T<T$_0$, $\rho/\rho_0 = 0$ up to 1T. However, for T$_{FM}$>T>T$_0$, the minimum resistivity value is reached at finite field. As we will show below, this resistivity minimum corresponds to the full in-plane saturation of the Eu moments, and we mark this field value as H$_{sat}$ (Fig.3B, black circles). In Figure 4, we plot H$_{sat}$ vs temperature and find that it is well described by square-root temperature dependence, $H_{sat} \propto \sqrt{T_{FM} - T}$, indicating the mean-field behavior of the Eu magnetic ordering. For 9K > T > 7.5K, we find that zero resistance can be induced in the vicinity of H$_{sat}$.

Following these measurements, sample 1 was mounted to a uniaxial stress device (see Methods; we corrected for a post-mounting background resistivity of order $\rho/\rho_0 = 1\%$, see Supp.Fig.S7). The sample was initially cooled under zero device voltage to base temperature. The sample was then slowly warmed under large fixed tension or compression, yielding the resistivity vs temperature curves in Figure 4 (right). The uniaxial stress was aligned along the Fe-As bonding direction, which induces strains in both B$_{1g}$ and A$_{1g}$ symmetry channels. We find that T$_{SC}$ varies monotonically as a function of the nominal uniaxial strain and can be tuned by ~1 K, revealing the tunability of the nematicity/superconductivity phase competition dominating by the A$_{1g}$ strain in line with previous work in BaFe$_2$As$_2$ (*25*, *26*). Intriguingly, the resistivity is especially tunable below T$_{FM}$, and we find that T$_0$ can be enhanced or suppressed by ~3 K, demonstrating the increased sensitivity to external tuning parameters within the ferromagnetic superconducting phase. However, T$_{FM}$ is virtually unchanged with strain, suggesting that strain has minimal impact on the Eu magnetic order.

Next, we applied field at fixed temperature and fixed tension or compression (Fig.3C). We find that the resistivity field dependence is similar to the zero-strain condition, but with an initial resistivity that is lower or higher with tension or compression, respectively, enabling a combined strain-field tuning of T$_0$. In Figure 4 we construct the strain and field-tunable phase diagram of superconductivity. Field-induced superconductivity is accessible in a temperature window from 7.5K to 9K under zero strain, with a measured minimum of 4K under compression and a maximum near 11K under tension, and with an onset field between μ₀H=0.1T and 0.3T. As the Eu magnetic moment increases with decreasing temperature, field-reorientation of the moments has a larger effect on the superconductivity, and so the fixed-strain phase volume of field-induced superconductivity is largest under compression and smallest under tension. We note that this is a substantial qualitative difference from UTe$_2$ where pressure tuning can shift the critical field by many tesla (*13*).

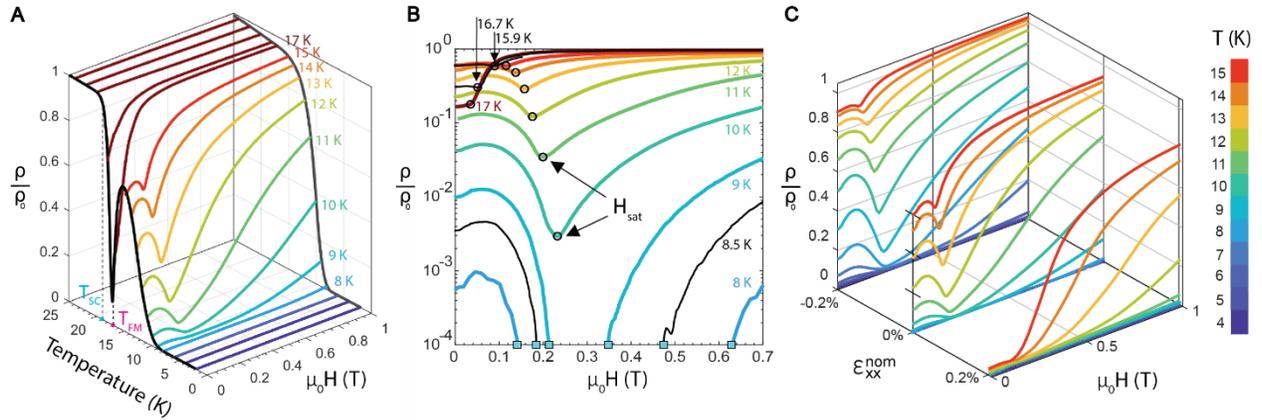

**Figure 3. Strain-tunable field-induced superconductivity.** (**A**) Freestanding resistivity vs temperature at fixed in-plane applied field ($\mu_0H=0$T, black; $\mu_0H=1$T, gray) and resistivity vs in-plane field at fixed temperature. Onset of superconducting transition ($T_{SC}=19$K) and ferromagnetic order ($T_{FM}=17.2$K) indicated. (**B**) Same resistivity vs field data as in (**A**) plotted against logarithm y-axis, with additional data at three non-integer temperatures (black). Cyan markers indicate entrance and exit from zero-resistance state for T=8K to 9K. Minimum of resistivity for T=10K to 16.7K and inflection point at 17K marked by black circles, corresponding to the in-plane saturation field $H_{sat}$ needed to align the Eu moments in-plane. Entrance and exit from zero-resistance state marked by cyan squares. (**C**) Resistivity vs field at fixed temperatures (4K to 5K) for one tensile and one compressive strain state and corresponding freestanding values from (**A**). Field range of zero-resistance shown in Fig.4 (shaded).

**Figure 4. Strain and Field Tunable Phase Diagram.** (Right) Resistivity vs temperature for the zero-strain state (same as black curve in Fig.2,3) and for the tensile (green) and compressive (magenta) strain states in Fig.3C. (Left) Phase boundary between $\rho>0$ and $\rho=0$ states under zero strain (cyan), tension (green) and compression (magenta), determined by resistivity vs temperature data (diamonds) and resistivity vs magnetic field (squares) from Fig.3 and Supp.Fig.S5. Field-induced superconductivity indicated by shaded areas for each strain state. Eu in-plane saturation field $H_{sat}$ taken from minimum of magnetoresistance in Fig.3B vs. temperature (black circles), with mean-field fit line (red).

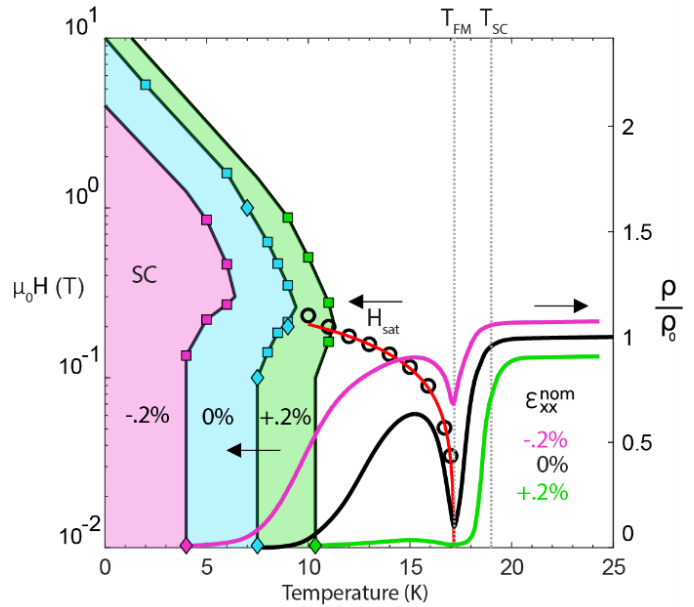

**Strain and Magnetic Field: Independent Tuning Knobs of Superconductivity**

To further identify the independence of strain and magnetic field for tunability of superconductivity, as well as to identify the mechanism of the field-induced superconductivity, we performed transport measurements under applied strain simultaneous with either x-ray diffraction (XRD) or x-ray magnetic circular dichroism (XMCD) at the Advanced Photon Source (see Methods). XMCD is a powerful tool to study ferromagnetic superconductors, as it is an element-specific electronic fluorescent effect which bypasses any diamagnetic shielding from the superconductivity. Further, it is a necessary tool for studying strain-tuning of the magnetic order given the experimental challenge of using conventional magnetometry techniques with a strain device (*27*).

We performed XRD measurements on sample 2 at T = 13.5 K, just below the maximum of the reentrant resistivity, across a range of strain. The linearity of the inline strain $\varepsilon_{xx}$ confirms a constant strain transmission (Fig.5B). We also measured the B$_{2g}$-symmetry spontaneous orthorhombicity $\varepsilon_S$, which is a proxy of the nematic order (*28*) (see Methods and Supp.Fig.S2). We find that under applied tension the magnitude of $\varepsilon_S$ is suppressed by up to 30%, coinciding with a dramatic decrease in the resistivity (Fig.5A). Under compression, $\varepsilon_S$ is roughly constant as the resistivity increases, suggesting that the saturated nematicity suppresses the superconductivity. This strain dependence of nematicity is consistent with the combination effect of the induced A$_{1g}$ and B$_{1g}$ strains, where the latter acts as a transverse field that suppresses nematicity quadratically. Thus, we can effectively strain-tune the superconductivity *via* its competition with the strain-tunable nematicity and the associated antiferromagnetic order (*26*, *29*).

Field-induced superconductivity was observed in sample 2 at T=10K under both fixed-strain and fixed-field conditions. With zero field, the resistivity can be strain-tuned from $\rho/\rho_0 = 5\%$ under zero strain, to $\rho/\rho_0 = 40\%$ at maximum compression, and $\rho/\rho_0 = 0\%$ with maximum tension (Fig.5C). Thus, tensile strain can effectively raise the superconducting transition to at least 10K. The application of an in-plane magnetic field (μ$_0$H=0.26 T) decreases the resistivity at all strain states, and zero resistivity is obtained at roughly 75% the maximum applied tension. Thus, tensile strain and magnetic field can work together to raise the transition temperature even higher. Figure 5D shows resistivity vs applied magnetic field at four fixed tension values, where a narrow strain range permits field-induced superconductivity.

To investigate the origin of the field-induced superconductivity, we performed simultaneous resistivity and XMCD measurements vs field at five fixed strain states between maximum compression and tension (Fig.5E,F). Here, the XMCD signal is proportional to the Eu magnetization along the field direction (10 degrees above the in-plane grazing incidence due to sample chamber constraints; see Methods). As the Eu ferromagnetic moments are spontaneously ordered along the c-axis, the Eu in-plane moment (and XMCD signal) is initially nearly zero under zero field. For all strains, increasing the magnetic field linearly increases the in-plane moment towards saturation at H$_{sat}$=0.25T, coinciding with the minimum of the magnetoresistance. From this, we conclude that the Eu moment reorientation towards the in-plane direction is intimately connected to field-induced superconductivity. Despite the large change in the zero-field resistivity with strain, there is no apparent strain-induced change in either the saturation field value or saturation XMCD value. This strain independence was somewhat unexpected given that the localized Eu 4f electrons presumably order with assistance from the strain-sensitive Fe 3d electrons via an RKKY interaction (*30*). As strain does not affect the Eu-magnetic order, and as strain has been shown to be far more effective than magnetic field in tuning the nematic order in this material system (*27*), we find that strain and field act as effectively independent tuning parameters of the superconductivity.

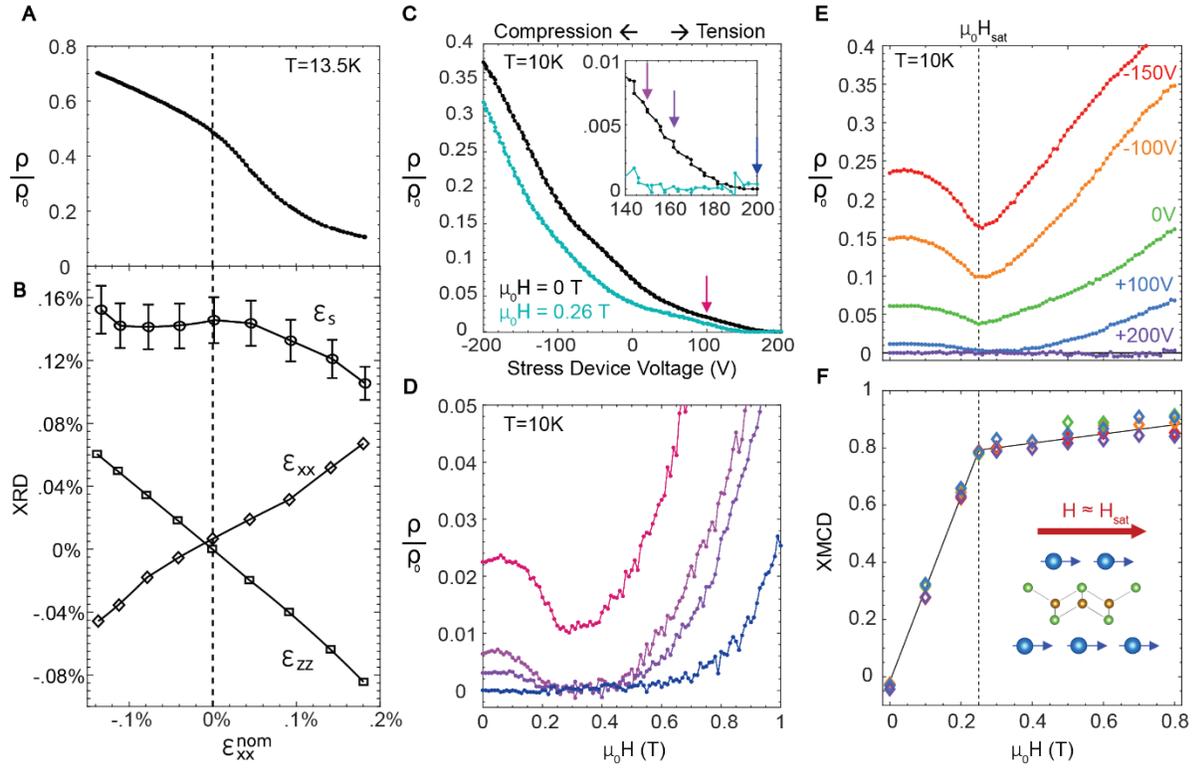

**Figure 5. X-ray characterization of independent strain and field tuning.** (**A**,**B**) Fixed temperature (T=13.5K) strain sweep (compressive to tensile) with simultaneous resistivity measurements (**A**) and XRD measurements (**B**) of the inline strain $\varepsilon_{xx}$, the out of plane strain $\varepsilon_{zz}$, and the nematicity-driven spontaneous orthorhombicity $\varepsilon_S$ (see Methods for definitions). (**C**) The resistivity vs strain device voltage at T = 10K under in-plane applied field of $\mu_0 H$=0T and $\mu_0 H$=0.26T. Inset shows high tension range. The voltage range in (**C**) corresponds approximately to the range of $\varepsilon_{xx}^{nom}$ in (**A**,**B**), which could not be simultaneously measured due to sample chamber restrictions. (**D**) Resistivity vs applied in-plane field at fixed strain values corresponding to colored arrows in (**c**) inset. (**E**,**F**) The simultaneously-collected resistivity (**E**) and XMCD (**F**) vs applied field at T=10K for five fixed strain values (see Methods for XMCD normalization details). Eu moment saturation coincides with minimum of resistivity at H=H$_{sat}$. Voltages listed in (**E**,**F**) correspond to slightly greater tension states than corresponding voltages in (**C**) due to different thermal hysteresis in the piezo actuators between the two measurements. Error bars in (**B**) on $\varepsilon_S$ represent error propagation of Gaussian fits to the split $[1\ 1\ 8]_T$ reflection peak (see Methods), while error bars on $\varepsilon_{xx}$ and $\varepsilon_{zz}$ are smaller than marker size.

**Mechanisms of field-induced superconductivity**

The Jaccarino-Peter effect (*31*) has often been invoked to explain field-induced superconductivity in s-wave superconductors, including in Eu-based Chevrel phases (*6*, *7*) and organic superconductors (*11*, *12*, *32*). Here, the Zeeman splitting induced by an external field compensates the internal exchange-bias splitting, resulting in superconductivity. In our case, the exchange-bias field is parallel to the external field, so a Jaccarino-Peter compensation is not possible. Instead, two other mechanisms contribute to the exchange splitting induced in the Fe bands: the Hund's rule coupling of Eu f- and d-orbitals, with the latter overlapping with Fe d-orbitals and inducing a polarization parallel to Eu f moments; and the Schrieffer-Wolfe coupling of Eu f- and Fe d- orbitals, which leads to an antiparallel polarization. To characterize these two effects, we performed DFT calculations using the Wien2K package (*33*, *34*) for Eu moments fully polarized in-plane (see Supplementary Information, Fig.S8-10). We find that both show high sensitivity to the Hubbard U on Eu sites, and as these two interactions have opposite signs, the induced splitting of Fe bands is relatively small and varying in sign and amplitude over the Fermi surface. This 'accidental cancelation' gives a reasonable explanation for the coexistence of superconductivity and ferromagnetism. Above $T_{FM}$ this cancellation is lifted as the Eu moments become disordered, which also explains the flipped field-preference of superconductivity above and below $T_{FM}$ (Fig.2). The small exchange interaction has also previously been suggested by DFT and Mossbauer studies in related materials (*18*, *35–37*). Nonetheless, these two exchange splitting mechanisms do not drive the field-induced superconductivity, and an alternative explanation is required.

An explanation to the field-induced superconductivity mechanism comes by considering the sizeable dipolar magnetic field exhorted by Eu moments onto the Fe layers. Using the classical Clausius-Mosotti theory of the polarizable media, we can determine the dipole field from the stacked infinite planes of fully-ordered ferromagnetic Eu moments as $B_{Eu} = m/3v = 0.3 T$, where $m = 7 \mu_B$ is the magnetization of the Eu moments and $v = 90$ Å$^3$ is the volume per moment. Importantly, this is not an "effective" magnetic field derived from the exchange splitting, but a real field (with respect to the superconducting condensate) that can be screened by Abrikosov vortices (*38*, *39*). Reviewing Figure 4F, at 10K and an applied field of 0.25T the XMCD signal saturates at 80% of the 2K XMCD value (Supp Fig. S4), suggesting a total dipole field of 0.24T, in agreement with this estimate. A resistive state is found under zero field, where a net 0.24T of Eu field is aligned to the c-axis. A zero-resistance state is found under an applied field of 0.25T in-plane, which combines with the reoriented Eu moments to give a total 0.49 T of flux in-plane. As in other iron-based superconductors (*40*), Eu(Fe$_{0.88}$Co$_{0.12}$)$_2$As$_2$ has a moderate in- vs. out-of-plane $H_{C2}$ anisotropy, with $\gamma = H_{C2,in}/H_{C2,out} \cong 2.1$ at T=2K (Supp.Fig.S5). As $\gamma > \frac{0.49\,T}{0.25\,T}$, and as we expect $\gamma$ to increase with temperature towards $T_{SC}$ (*41*), we can explain the narrow field range of the field-induced superconductivity as due primarily to rotating the Eu moments in-plane to take advantage of the higher in-plane critical field. Further, this explains why applied strain does not shift the field range where superconductivity onsets, as strain does not directly tune the Eu magnetic order.

**Discussion**

Here, we have demonstrated field-induced superconductivity at T=9K, which can be accessed at small field (μ$_0$H≤0.3T) and tuned with accessible strain values (|ε$_{xx}$|<0.2%). Our combined XRD, XMCD and transport measurements show that strain and magnetic field act as independent tuning knobs, with the former affecting the nematic order and Fe antiferromagnetism and the latter affecting the Eu ferromagnetism. These knobs tune the phase diagram analogously to chemical doping, but without introducing additional disorder. The high tunability of this system results from the close competition between the simultaneously coexisting superconducting, nematic and ferromagnetic phases. In contrast, no field-induced superconductivity has been reported in related Eu-based iron pnictide materials such as EuRbFe$_4$As$_4$ (*42*) or optimal Ir-doped EuFe$_2$As$_2$ (*15*), likely due to the substantially stronger superconducting order. We anticipate even higher field-induced superconducting temperatures could be obtained in materials engineered with a perfect balance between higher temperature superconductivity and ferromagnetism.

An open question is the microscopic details of the zero-resistance state, and especially its strain-tunability. The small resistivity just above T$_0$ has previously been associated with mobile flux vortices making up a spontaneous vortex liquid phase, with zero resistivity indicating the freezing of these vortices (*2*). An intriguing possibility to explain the enhanced strain-tunability of T$_0$ below T$_{FM}$ is that vortices become pinned at nematic domain boundaries (*43*), which can be tuned in number and size with strain. A second direction for future work is to assess this material's potential for superconducting spintronics applications, such as by studying the degree of spin polarization and spin-triplet pairing of the supercurrent as it passes through the field-tunable magnetic layers (*4*).

We have also described a novel mechanism for field-induced superconductivity distinct from the Jaccarino-Peter effect and spin-triplet U-based compounds. This mechanism could likely be present in other systems that exhibit (1) large magnetic moments that are easily field-tunable (e.g. L=0 rare earth elements) and (2) a superconducting order which is dimensionally highly anisotropic (e.g. a van der Waals (vdW) material (*44–46*) or at the interface between different materials (*47*)). This mechanism could arise quite naturally from a vdW heterostructure, with one superconducting layer and one ferromagnetic layer. We note that the apparent first report of field-reentrant superconductivity in a vdW system occurs with stacked thin flakes of antiferromagnetic CrCl3 and superconducting NbSe$_2$ (*48*), which demonstrates the potential for our proposed mechanism to likewise underlie field-induced superconductivity in 2D materials.

**Acknowledgements:** We thank Riccardo Comin, Connor Occhialini, Lukas Powalla, Xueqiao Wang, SY Frank Zhao, Alan Chen, Caolan John, Jagadeesh Moodera and others for useful conversation.

**Funding:**

Air Force Office of Scientific Research under grant FA9550-21-1-0068 and the David and Lucile Packard Foundation (J.J.S., J.M.D., E.R., Y.S., P.M., and J-H.C.)

U.S. Department of Energy (DOE), Office of Science, and Office of Basic Energy Sciences, Contract No. DE-AC02-06CH11357 (G.F., Y.C., J.W.K., and P.R.)

U.S. Department of Energy (DOE), grant No. DE-SC0021089 (I.I.M)

National Natural Science Foundation of China, Grant No.11904319. (Y.H.)

National Science Foundation MPS-Ascend Postdoctoral Research Fellowship, Award No. 2138167 (J.J.S.) Any opinions, findings, and conclusions or recommendations expressed in this material are those of the author(s) and do not necessarily reflect the views of the National Science Foundation.

**Author Contributions:**

J.J.S conceived the project and wrote the paper, with input from all authors. J.J.S. and Y.S. grew the samples. J.J.S., J.M.D., E.R., Y.S., and P.M. performed all non-synchrotron transport measurements. J.J.S., G.F., Y.C., J.W.K., and P.R. performed all synchrotron measurements. The field-induced superconductivity mechanism proposed in this paper was conceived by J.J.S. and further developed by I.I.M., while Y.H. and I.I.M. performed DFT calculations. J.H.C. and P.R. supervised the project.

**Competing Interests Statement:** The authors declare no competing interests.

**Data Availability:** The data supporting the findings in this paper are available from the corresponding author on request.

**Supplementary Materials**

Materials and Methods

Supplementary Text

Figs. S1 to S10

# Supplementary Materials for

# Strain-Switchable Field-Induced Superconductivity

**Authors:** Joshua J. Sanchez[1,3,*], Gilberto Fabbris[2], Yongseong Choi[2], Jonathan M. DeStefano[3], Elliott Rosenberg[3], Yue Shi[3], Paul Malinowski[3,4], Yina Huang[5], Igor I. Mazin[6], Jong-Woo Kim[2], Jiun-Haw Chu[3,*], Philip Ryan[2,*]

[1]Department of Physics, Massachusetts Institute of Technology, Cambridge, MA 02139, USA

[2]Advanced Photon Source, Argonne National Laboratories, Lemont, IL, USA.

[3]Department of Physics, University of Washington, Seattle, WA, USA.

[4]Department of Physics, Cornell University, Ithaca, NY 14853, USA.

[5]School of Science, Zhejiang University of Science and Technology, Hangzhou 310023, China

[6]Department of Physics and Astronomy and Quantum Science and Engineering Center, George Mason University, Fairfax, Virginia 22030, USA

*Corresponding author: jjsanchez.physics@gmail.com, jhchu@uw.edu, pryan@anl.gov

**The PDF file includes:**

**Materials and Methods**

**I. Crystal Growth**

**II. XRD under strain**

**III. XMCD of 2K result**

**IV. Freestanding magnetoresistance Sample 1**

**V. Assessing the nonzero background resistance of Sample 1**

**VI. Density Functional Theory Analysis of Eu-Fe exchange interactions**

**Figs. S1 to S10**

**Materials and Methods**

**Sample Preparation**

Single crystal samples of $\mathrm{Eu(Fe_{0.88}Co_{0.12})_2As_2}$ were grown from a tin flux as described elsewhere (*23*). We used a nonstoichiometric mix ratio of Eu: $\mathrm{(Fe_{0.85}Co_{0.15})}$: As: Sn of 1:8.5:2:19. This ratio resulted in samples with higher zero-resistance temperatures ($T_0$) compared to the stoichiometric 1:2:2:20 ratio (*23*), (Supp.Fig.S1). However, there was significant sample to sample variability in $T_0$, which may result from doping inhomogeneity. The composition was measured by EDX to be 12% Co-doping, despite a nominal doping of 15%. The samples were cleaved from large as-grown single crystal plate and cut along the tetragonal [1 0 0] direction into bars with dimensions ~2 x 0.60 x 0.06 mm. Four gold wires were attached with silver epoxy to measure the inline resistivity $\rho_{xx}$ using a standard 4-point measurement and an SR830 lock-in amplifier with 1mA fixed current. Sample 1 was measured in a Quantum Design PPMS. Sample 2 was measured in x-ray compatible cryostats at Argonne National Laboratory.

A piezo-actuator uniaxial stress device (Razorbill Instruments, CS-100) was used to provide in-situ stress. The built-in capacitance strain gauge was used to determine the nominal strain $\varepsilon_{xx}^{nom}$ as in (*28*). Sample chamber constraints prevented the measurement of $\varepsilon_{xx}^{nom}$ for data presented in Figure 5C-F. Below the nematic transition (Ts=68K, Supp.Fig.S3), structural twin domains form along the Fe-Fe bonding direction ([1 1 0]$_T$, with lattice constants $a_{or}$ and $b_{or}$), with orthorhombicity $\varepsilon_S = \frac{a_{or}-b_{or}}{a_{or}+b_{or}}$. In this work, we apply stress along the Fe-As bonding direction ([1 0 0]$_T$, with lattice constant $a_T$), resulting in an inline strain $\varepsilon_{xx} = \frac{\Delta a_T}{a_{T,0}}$ and an out of plane strain $\varepsilon_{zz} = \frac{\Delta c}{c_0}$. The applied stress thus does not detwin the domains, but instead can tune the magnitude of the nematic order parameter through nonlinear couplings between $\varepsilon_{xx}$, $\varepsilon_{zz}$ and $\varepsilon_S$ (see ref. (*25*) and Supp. Figs. 2-3).

After mounting sample 1 on the strain device, a field, strain and temperature dependent background resistivity of order $\rho/\rho_0 \approx 1\%$ was present, masking the true entrance into the zero-resistance state. We estimate the field range of field-induced superconductivity from the field range where the resistivity dips below this background resistivity (see Supp Fig.S6,7 for analysis), from which we estimate field-induced superconductivity occurs in the bulk of the sample up to T=11K. True zero resistance was measured in sample 2 at T= 10K, reported in Fig.5C,D.

**X-ray Magnetic Circular Dichroism and X-ray Diffraction**

XRD measurements were performed at the Advanced Photon Source, beamline 6-ID-B, at Argonne National Laboratory. X-rays of energy 7.6 keV illuminated an area 500x500 um, fully encompassing a cross section of the middle of the crystal where strain transmission is highest. The sample and strain device were mounted on a closed cycle cryostat. Gaussian fits to the tetragonal (1 0 7), (0 0 8) and (1 1 8) reflections were used to determine the lattice constants ($a_T$), (c), and ($a_{Or}$ & $b_{Or}$), corresponding to in-plane along the stress axis, out of plane, and in-plane at 45 degrees to the stress axis, respectively.

XMCD was measured at the Advanced Photon Source beamline 4-ID-D at Argonne National Laboratory. We probed the Eu L$_3$ edge using x-rays of 6.97 keV, which measure the spin polarization of the Eu 5d band due primarily to the magnetic moment of the 4f orbital. A superconducting split coil magnet with a large bore was used to apply magnetic field. The sample temperature was controlled using He flow. XMCD was collected in fluorescence geometry by monitoring the Eu $L_\alpha$ line using a four element Vortex detector integrated with the Xspress module to enable a larger dynamical range. Circularly polarized x-rays were generated using a 180 microns thick diamond (111) phase plate. Data was corrected for self-absorption. The XMCD spot size illuminates the whole sample width across the y direction and is roughly 100 microns wide along the x direction (between the transport wires) and probes a depth of about 5 microns. The beam is centered on the middle of the crystal where strain is most transmitted and homogenous. The incident beam was aligned with the applied magnetic field at an angle of ~10 degrees above parallel to the sample surface (grazing incidence) due to sample chamber constraints. All XMCD data is normalized to the zero-strain, $\mu_0 H > 0.3$T saturated value at T=2K (Supp. Fig.S4).

**DFT Calculations**

The full-potential linearized augmented plane wave Wien2K package (*33*) has been used for the DFT calculations. We use the Perdew, Burke, and Ernzerhof (*34*) version of the generalized gradient approximation (GGA) to the exchange-correlation functional within density functional theory. The sphere radii for Eu, Fe, As are taken as 2.50, 2.29, 2.18 bohr, respectively. The basis set cut-off parameter RmtKmax = 8.0 was used. The number of k points was set to 4500. The crystal structure and magnetic moments on Eu and Fe are illustrated in Fig. S8. We set U=9eV on the Eu atom and did collinear spin polarized self-consistent calculations in the primitive (not conventional) cell. WIEN2k has a parameter ($\kappa$) which tweaks the strength of Hund's rule coupling. The Hund's Rules coupling is set to normal full strength when $\kappa=1$ and completely switches off when $\kappa=0$. We used this parameter to delineate the two effects mentioned above: the Schrieffer-Wolfe interaction does not depend on the Hund's rule coupling strength, while the Eu(f)-Eu(d) interaction can be switched off using $\kappa$. In Fig. S9, we show the band structure for these two values of $\kappa$ around the Fermi level. The largest splitting near the Fermi level when $\kappa=1$ is about 25meV along $\Gamma$Z-.

**Supplementary Text**

**I. Crystal Growth**

Single crystal samples of $\text{Eu}(\text{Fe}_{0.85}\text{Co}_{0.15})_2\text{As}_2$ were grown from a tin flux as described elsewhere (*23*). We used a nonstoichiometric mix ratio of Eu: $(\text{Fe}_{0.85}\text{Co}_{0.15})$: As: Sn of 1:8.5:2:19. This ratio resulted in samples with higher zero-resistance temperatures ($T_0$) compared to the stoichiometric 1:2:2:20 ratio (Fig.S1). However, EDX measurements suggest the Co-doping to be 12% instead of the nominal 15%. The superconducting properties of doped EuFe$_2$As$_2$ depend sensitively on growth methods. For instance, no superconductivity is found for Co-doped samples grown by FeAs flux (*49*). Further investigation of the non-stoichiometric growth conditions is warranted.

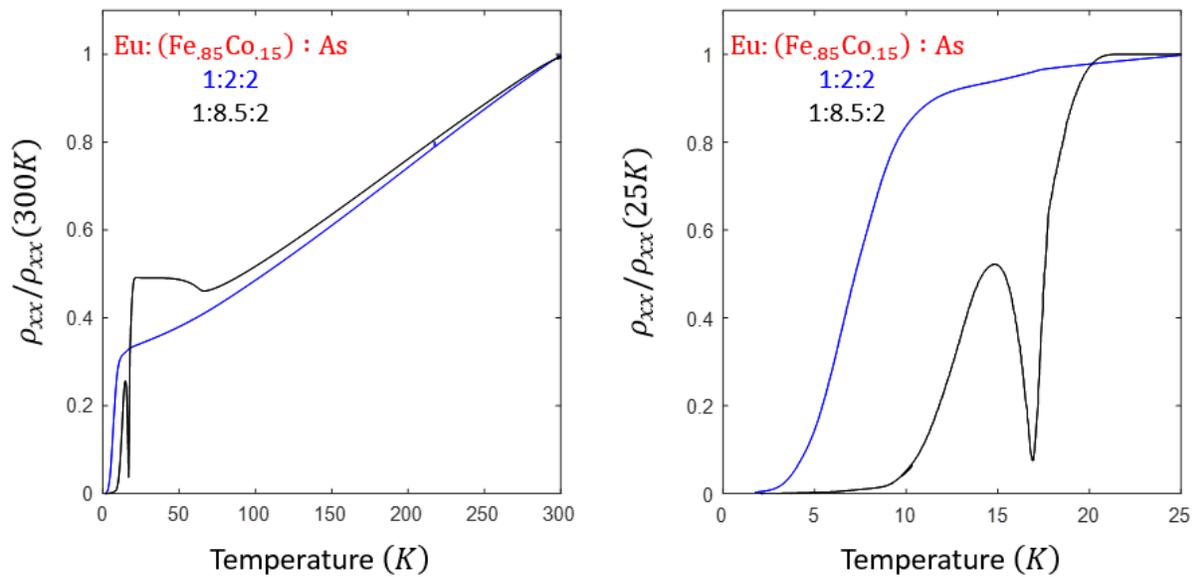

**Figure S1.** Comparison of resistivity vs temperature for samples grown with an elemental composition that was stoichiometric (blue) and nonstoichiometric (black; sample 2 from main text).

**II. XRD under strain**

XRD measurements were performed on sample 2 at the Advanced Photon Source, beamline 6-ID-B, at Argonne National Laboratory. X-rays of energy 7.6 keV illuminated an area 500x500 um, fully encompassing a cross section of the middle of the crystal where strain transmission is highest. The sample and strain device were mounted on a closed cycle cryostat. Gaussian fits to the tetragonal (1 0 7), (0 0 8) and (1 1 8) reflections were used to determine the lattice constants ($a_T$), (c), and ($a_{Or}$ & $b_{Or}$, due to the split peak in the twinned state), corresponding to in-plane along the stress axis, out of plane, and in-plane at 45 degrees to the stress axis, respectively.

Figure S2 shows the uniaxial strains $\frac{\Delta a_T}{a_T}$ and $\frac{\Delta c_T}{c_T}$ and the nematic-driven spontaneous orthorhombicity $\varepsilon_S = \frac{a_{Or}-b_{Or}}{a_{Or}+b_{Or}}$ as a function of strain at T=13.5 K. We observe that $\varepsilon_{xx} = \frac{\Delta a_T}{a_{T,0}}$ is roughly linear to $\varepsilon_{xx}^{nom}$ determined from the capacitive strain gauge of the stress device. We find a nearly constant strain transmission of $\frac{d\varepsilon_{xx}}{d\varepsilon_{xx}^{nom}}$ =34% across the range of the strains applied. The out of plane uniaxial strain $\varepsilon_{zz} = \frac{\Delta c}{c_0}$ is surprisingly large compared to $\varepsilon_{xx}$ with the Poisson ratio $\nu_{xz} = -\frac{d\varepsilon_{zz}}{d\varepsilon_{xx}}|_{\varepsilon_{xx}=0}$ approaching 1.3. A Poisson's ratio greater than unity is unexpected in an isotropic, linear elastic material, and so this large change in planar spacing may result from a significant magnetostructural response due to the ferromagnetic Eu layers, as well as tuning of the nematic order.

Most significantly, $\varepsilon_S$ is found to be suppressed by roughly 30% at maximum tension, while being relatively unaffected (or even slightly enhanced) by compression. Given the competition between nematic order and superconductivity, it is clear that the sharp reduction in the resistivity with tension can be attributed (at least in part) by a strain-suppression of the nematic order (see also Fig.S6a). This result is fully in agreement with previous work in Co-doped BaFe$_2$As$_2$, where tension (compression) applied along the tetragonal [1 0 0] direction resulted in a suppression (enhancement) of the nematic transition temperature (*25*). We observe a similar effect via warming the sample through the nematic transition under either tension or compression (Figure S3). Phenomenologically, the tuning of nematicity with stress applied along the tetragonal [1 0 0] direction results from both to the introduction of an orthogonal antisymmetric strain $\left(\varepsilon_{B_{1g}} = \frac{a_T-b_T}{a_T+b_T}\right)$ which acts to suppress nematicity and $\varepsilon_S$, and to a surprisingly large sensitivity to $\varepsilon_{zz}$ which tunes the unit cell volume despite not breaking any symmetries (see ref. (*25*)).

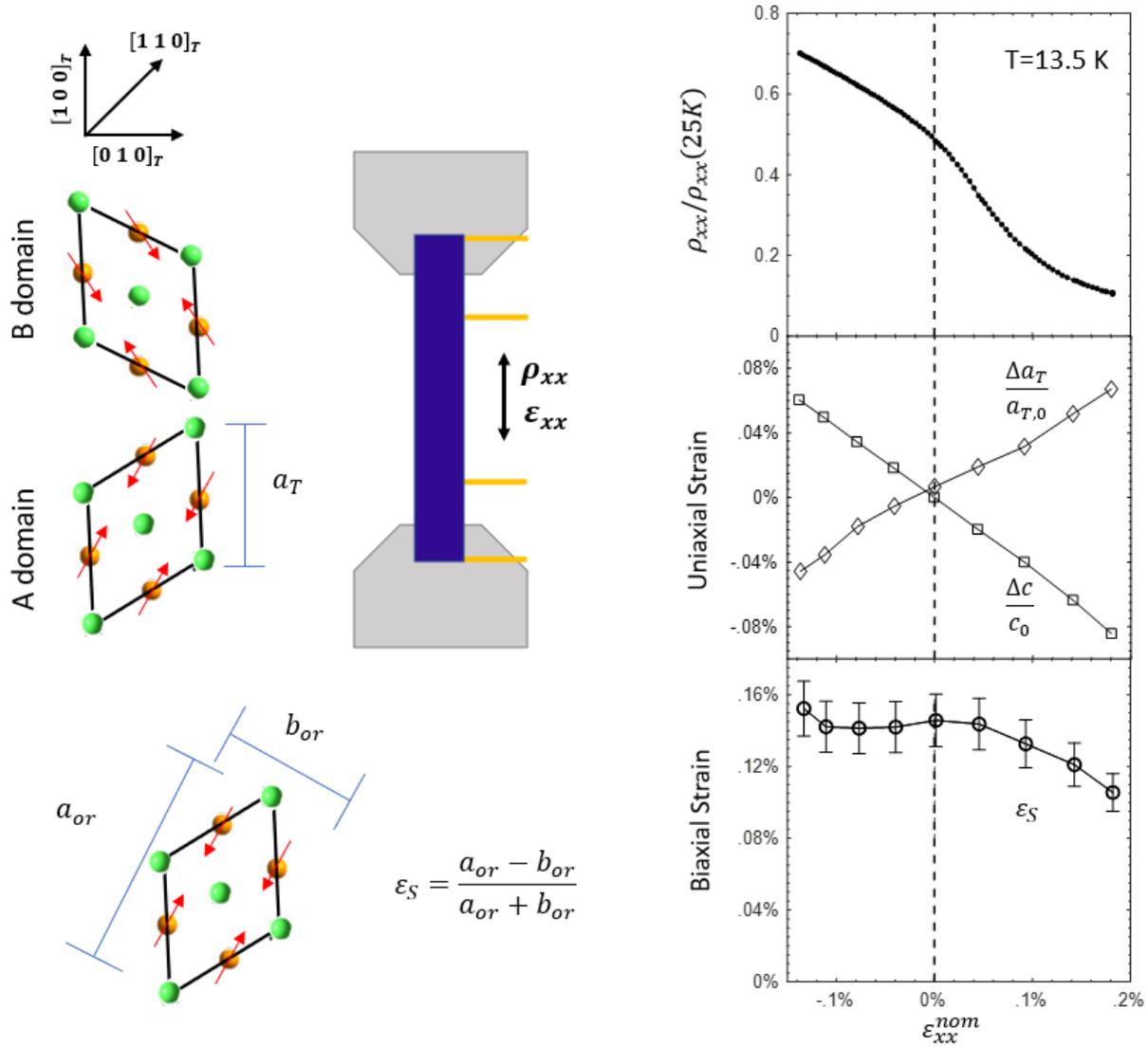

**Figure S2.** Sample 2. Fixed temperature (T=13.5K) strain sweep (compressive to tensile) with simultaneous resistivity measurements and XRD measurements of the $a_T$ and $c$ lattice constants (presented normalized by their zero-strain values) and the $a_{or}$ and $b_{or}$ lattice constants (presented as the antisymmetric strain $\varepsilon_{B_{2g}}$). Error bars on $\varepsilon_{B_{2g}}$ represent the error propagation of the Gaussian fits to $a_{or}$ and $b_{or}$.

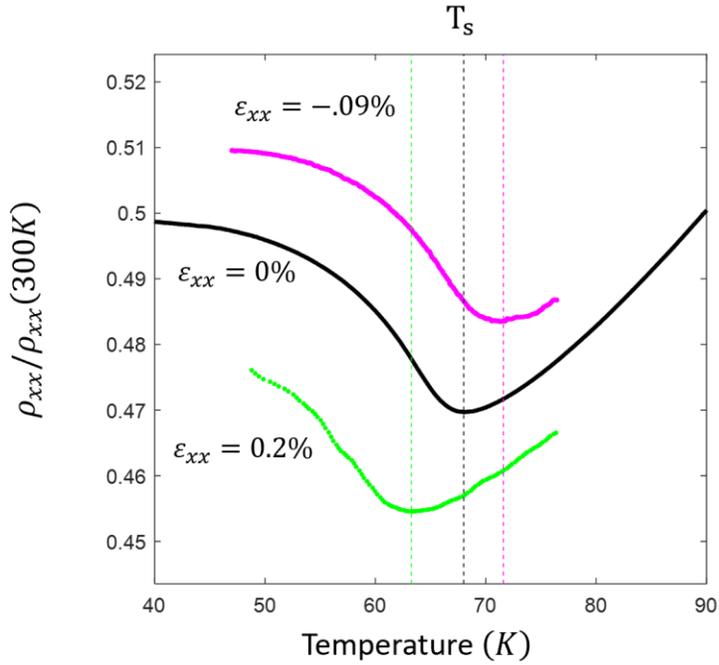

**Figure S3.** Sample 2. Resistivity vs temperature under zero strain (black, freestanding data from Fig.S1), tension (green) and compression (magenta). Note: the fixed-strain data has been corrected for a ~3 K thermal lag. As such, we do not attempt to make a quantitative assessment of the strain-tuning of the transition temperature, and instead only share this data to show the basic phenomenology of an enhanced (suppressed) nematic transition temperature with compression (tension) in qualitative agreement with past work in Co-doped $BaFe_2As_2$ (*25*).

### III. XMCD of 2K result

On sample 2, the first XMCD data were taken after the initial cooldown at zero applied strain at T=2 K and 10 K through a field range of $\mu_0 H = \pm 1$ T. All XMCD data in this work are normalized to this $\mu_0 H > 0.3$ T, T= 2 K fully saturated XMCD value, which corresponds to the $M \sim 7\ \mu_B$ ($\mu_0 H \sim 0.3$ T) fully ordered Eu magnetic moment. The initial XMCD saturation value at T=10K, $\mu_0 H$=0.25T is approximately 80% of the 2K saturation value. See Main Text for details of the XMCD measurement.

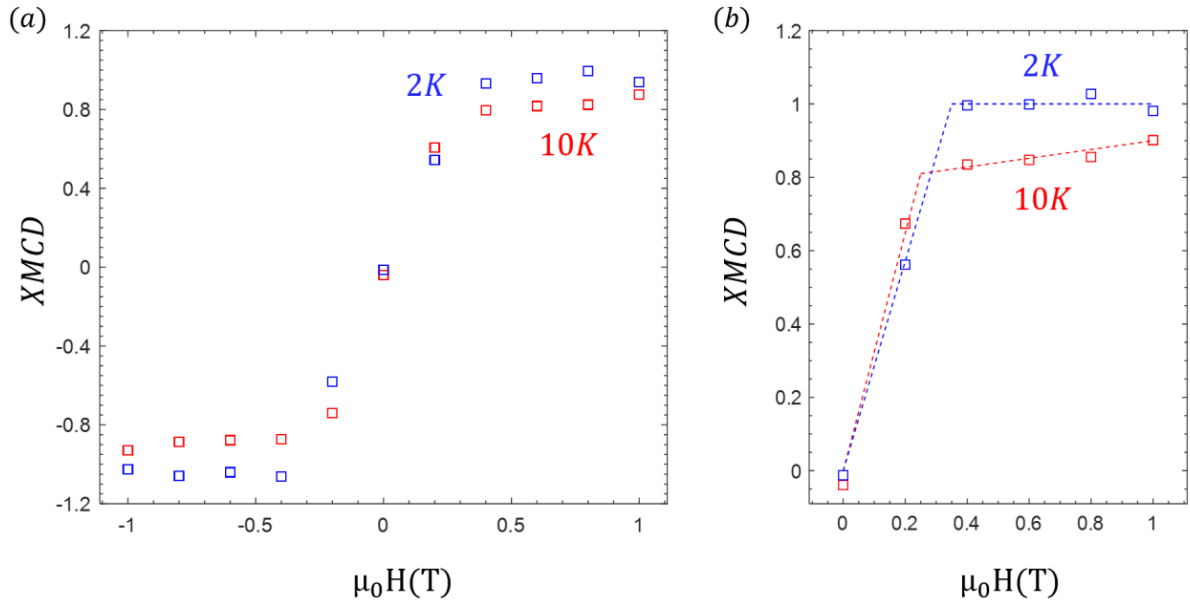

**Figure S4.** XMCD vs field at 2K and 10K. Data in (a) collected from a single field sweep from +1T to -1T. Data in (b) are the normalized difference of the positive and negative field values in (a).

## IV. Freestanding magnetoresistance Sample 1

In the freestanding state prior to mounting on the strain cell, sample 1 was cooled through the superconducting and ferromagnetic transitions under zero field (Fig.S5a,b, black), and with an applied field of µ₀H=0.1 T, 0.2 T and 1.0 T either in-plane (Fig.S5a) or out of plane (Fig.S5b). An out of plane field is found to only increase the resistivity, while only lowering the value of $T_0$. In sharp contrast, an in-plane field is far more detrimental to superconductivity between $T_{sc}$ and $T_{FM}$, but zero resistance is reached at an enhanced value of $T_0$=9.0 K for µ₀H=0.2 T, demonstrating field-induced superconductivity.

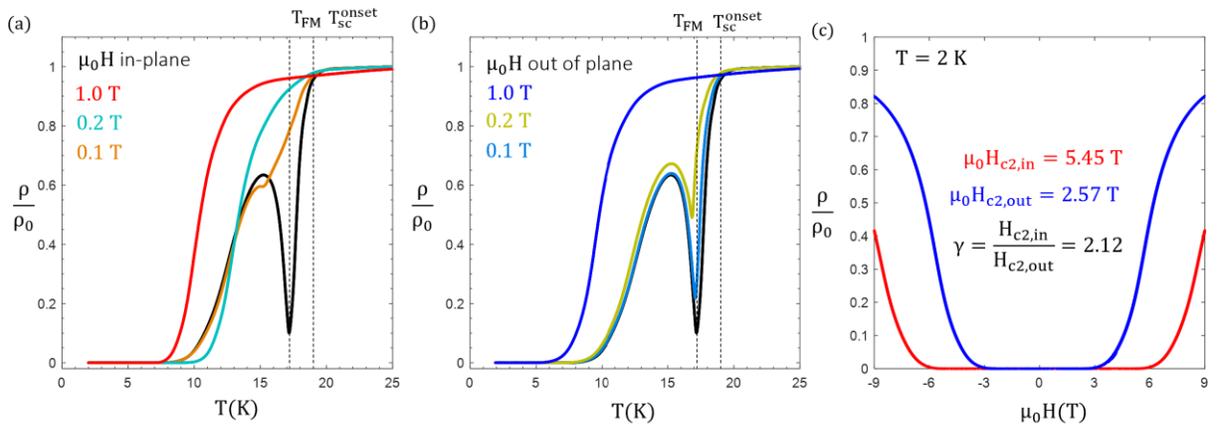

**Figure S5.**

**(a,b)** Sample 1 resistivity vs temperature for zero applied field (black) and $B = 0.1\,T, 0.2\,T$ and $1\,T$ applied in-plane **(a)** and out of plane **(b)**. For $\mu 0H = 0.2\,T$ applied in-plane, the zero-resistivity temperature rises from $T_0 = 7.5\,K$ to $9.0\,K$. **(c)** At T=2 K magnetic field was applied in plane (red) and out of plane (blue) to extract the upper critical fields $H_{c2}$ for each direction, yielding an anisotropy term $\gamma \cong 2.1$.

## V. Assessing the nonzero background resistance of Sample 1

Extensive resistivity measurements of sample 1 were made under different temperature, field and strain states. Unfortunately, after mounting sample 1 on the strain device, a field, strain and temperature dependent background resistivity of order $\rho \approx 0.01\%$ was present, masking the true entrance into the zero-resistance state (Fig.S6). This is apparently due to a small volume of the sample which buckled under strain and thus behaves as if heavily-compressed, effectively raising its respective value of $T_0$ while still being highly strain, field and temperature dependent.

Here we describe the workaround to this issue. Prior to mounting on the strain device, the sample reached zero resistance at $T_0 = 7.5\ K$. In Figure S6, the orange trace is under small tension ($\varepsilon_{xx} = 0.04\%$) and has a lower resistivity than the freestanding trace at all temperatures above T=9K, indicating an enhancement to the superconductivity. Below T=9 K, the orange trace has a higher resistivity and never reaches zero. The sample should be expected to reach zero resistance at higher temperature under tension, as is observed in sample 2. At T=7.5K, the orange trace should already be in a zero-resistance state, but instead has a value of approximately $\frac{\rho}{\rho_0} = 1\%$. We thus use this value as a conservative estimate for the temperature and field entrance into the true zero resistance state. We determine the value of $T_0$ under fixed strain as $T_0(\varepsilon_{xx} = -.19\%) = 4K$, $T_0(\varepsilon_{xx} = .04\%) = 7.5K$, and $T_0(\varepsilon_{xx} = .20\%) = 10.3K$.

In Figure S7 we show the resistivity vs applied in-plane magnetic field data at several temperatures under $\varepsilon_{xx} = -.19\%$ (a) and $\varepsilon_{xx} = .20\%$ (b). The field values where the resistivity crosses $\frac{\rho}{\rho_0} = 0.01\%$ are indicated by square markers, and these values are used to define the field-induced superconductivity phase space in the Main Text Figure 4. Note that the resistivity below this cutoff value appears to still show a temperature, field, and strain tunability, but this occurs separately from the tunability of the bulk of the sample.

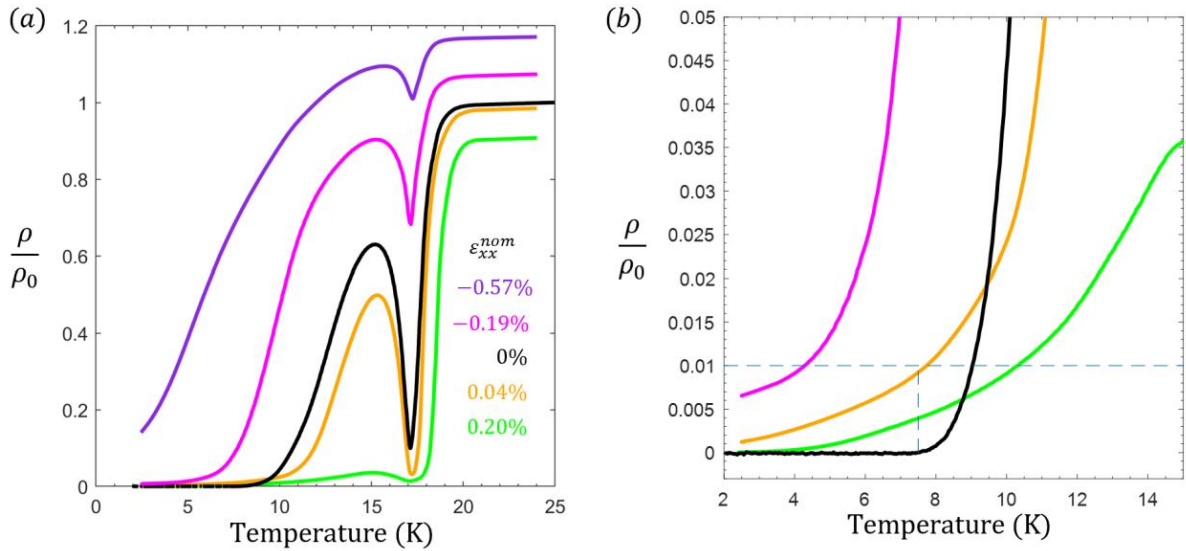

**Figure S6. (a)** Resistivity vs temperature at two compressive and two tensile strain values, compared to freestanding value (black). **(b)** The same data zoomed to observe low resistivity values.

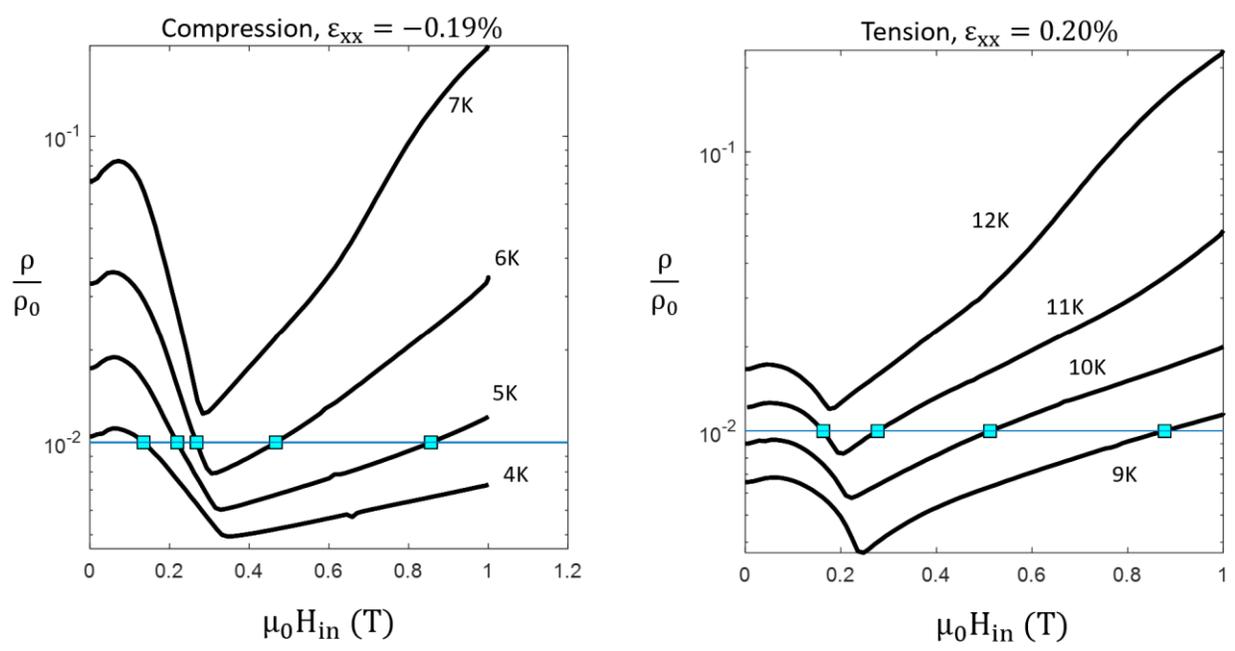

**Figure S7.** Sample 1. Resistivity vs applied in-plane magnetic field under compression (a) and tension (b). $\rho/\rho_0$>0.01 line marks the cutoff for estimated entrance into bulk zero-resistance state.

**VI. Density Functional Theory Analysis of Eu-Fe exchange interactions**

A puzzle is presented by the finding that an in-plane field is far more detrimental to superconductivity than an out of plane field above the Eu ferromagnetic ordering ($T_{SC} > T > T_{FM}$), while it promotes superconductivity below the maximum of the reentrant resistivity (Supp. Fig.S5). A plausible explanation can be given by considering the two competing effects of Eu moments on the Fe spin polarization. First, when Eu moments are aligned in-plane, the direct hybridization between Fe 3d states with Eu 4f states creates a spin-polarizing exchange bias, which pushes Eu-aligned Fe spin states at the Fermi surface up in energy and anti-aligned states down (Schrieffer-Wolfe interaction). This induces a direct exchange splitting $\Delta_d \sim t_{Fe(d)-Eu(f)}^2/(E_F - E_{Eu(f)})$, where the induced Fe polarization is *antiparallel* to that of the Eu ions. Second, there is a competing interaction due to hybridization between the Fe bands and the empty Eu d-bands. Since the latter are coupled to Eu 4f states though Hund rule's coupling, this effect leads to a net polarization of Fe that is *parallel* to Eu moments. A net polarization of Fe moments in *either* direction suppresses Cooper pairing.

In order to estimate these effects quantitatively, we have performed DFT calculations using the Wien2K package (*33*, *34*) to characterize these two effects. We find that both show high sensitivity to the Hubbard U on Eu sites, and as both are small and opposite in sign, their sum can be of either sign, or even of different signs in different bands. Below the Eu ferromagnetic transition, these two effects seem to largely cancel, and the dipole-field mechanism is dominant. Above the transition, the Eu moments do not fully saturate at small applied field, and the (accidental) cancelation between the two electronic effects will be lifted. This causes the itinerant Fe moments to gain a net magnetization for Eu moments aligned in-plane, which yields the stronger suppression of superconductivity with in-plane field. While we cannot calculate this exchange field with the required accuracy, given the strong dependence on the applied Hubbard correction, we can estimate the order of magnitude of the effect. We find that for fully in-plane polarized Eu moments (with $m = 7\ u_B$) and a variable U, the exchange interaction is within ±10 meV. This is far larger than the anticipated superconducting gap of the order of 2 meV (given $T_{SC}$ = 19K), and so the scale of the effect – barring the above-discussed accidental cancellation – is on the right order to suppress superconductivity for the in-plane field. Meanwhile, for an out-of-plane applied field, Eu and Fe moments are perpendicular and the exchange splitting is only second order in the exchange field, resulting in a much weaker Eu-driven polarization of Fe spins and a far milder suppression of superconductivity. This explains why the effect of magnetic field shows the opposite anisotropy above and below the ferromagnetic ordering temperature of Eu.

Here we discuss the details of our calculations. We have used density functional theory (DFT) to characterize the effects of the Eu moments on the antiferromagnetically-ordered Fe moments via different exchange interactions, in order to estimate the band effects of Eu magnetism on Fe-origin superconductivity. We directly observe that both parallel and antiparallel Eu-Fe alignment effects are present, which appear to largely cancel out at low temperature.

We also show projected Fe d, Eu f and Eu d bands for κ=0 in Fig. S10. In the κ=0 case, Fe1 and Fe2 have equivalent small moments that have the same direction as the Eu moment, so we do not differentiate Fe1 and Fe2 in projected band. We can clearly see Fe d spin up bands pushed up above the Eu f bands, and this effect is larger closer to Eu f bands. Another effect is a non-negligible Fe d spin up contribution in the Eu f bands.

Comparing calculations with the full Hund's rule coupling ($\kappa=1$), where both Schrieffer-Wolfe Eu(f)-Fe(d) antiferromagnetic coupling and Hund's rule Eu(f)-Eu(d) ferromagnetic coupling is included, and those with $\kappa=0$, where only the former is operative, we see that (a) Eu(f)-Fe(d) is indeed antiferromagnetic (the red bands are always below the blue ones), and rather large for some bands at the Fermi level, while the competing Eu(f)-Eu(d) interactions is largely cancelling it for $\kappa=1$, and this cancellation is, fortuitously, nearly complete right at the Fermi level (while at ~0.2 eV below or above it becomes large, up to 100 meV). This confirms our conjecture of a fragile character of this cancellation.

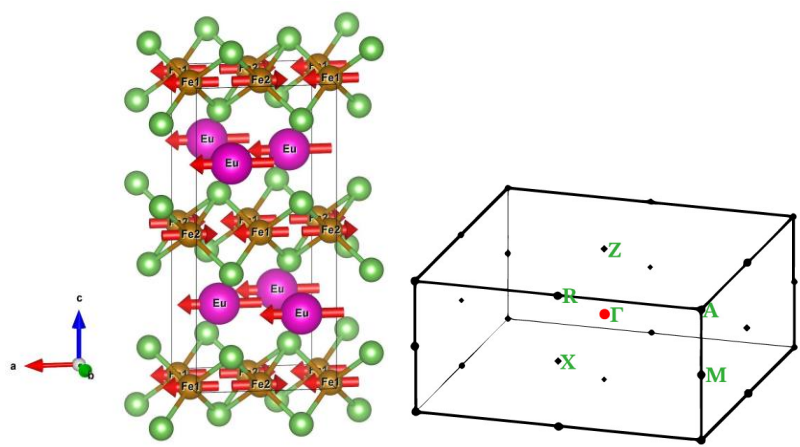

**Figure S8.** Conventional structure, with a= 5.5372 Å, b= 5.5052 Å, c= 12.0572 Å. Eu atoms are ferromagnetically ordered along the easy axis of the Fe antiferromagnetic order. High symmetry points of corresponding BZ are shown on right.

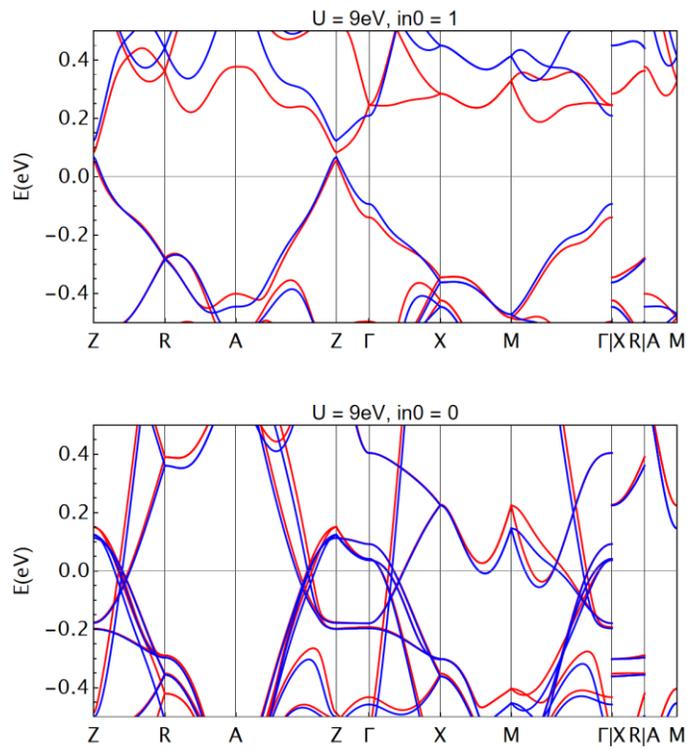

**Figure S9.** Band structure of U=9eV for κ=0 and 1. Red bands for spin up (aligned to Eu moments) and blue bands for spin down (anti-aligned to Eu moments).

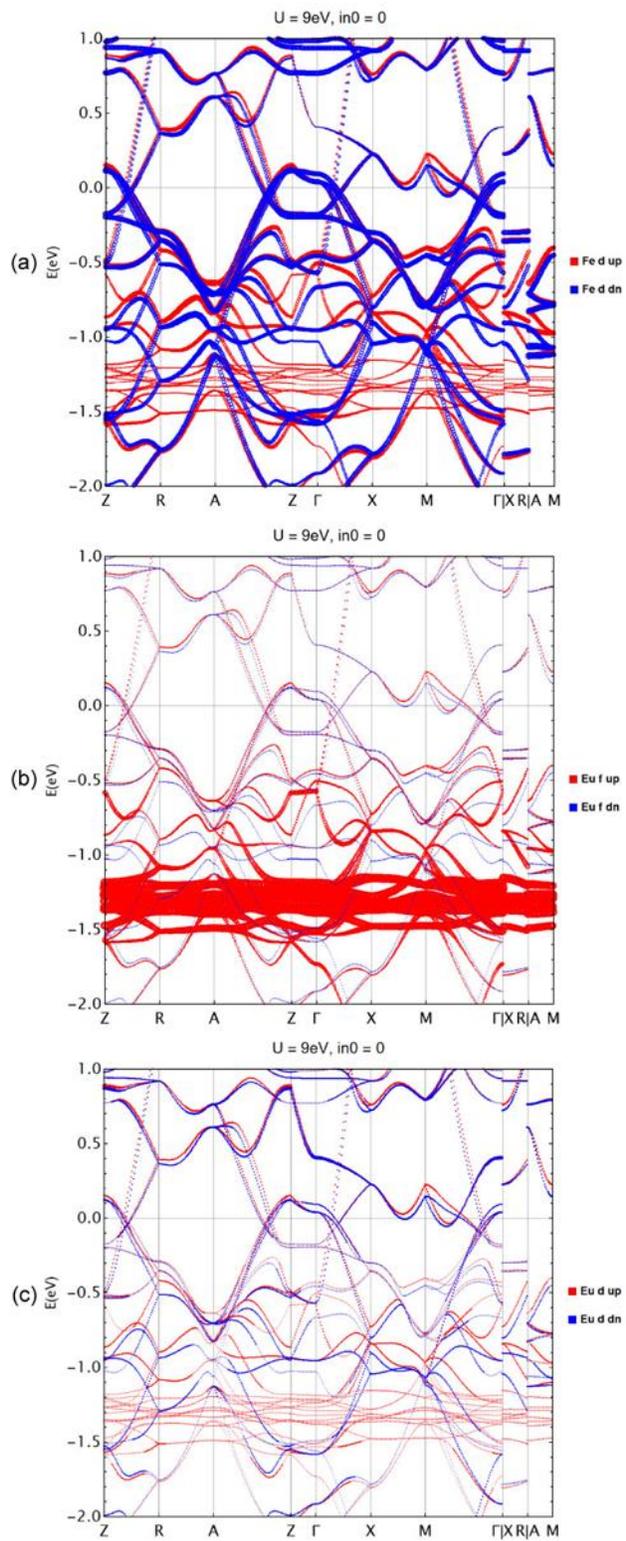

**Figure S10.** Projected bands of (a) Fe d, (b) Eu f and (c) Eu d.